\documentclass[submission, Phys]{SciPost}

\hypersetup{
    colorlinks,
    linkcolor={red!50!black},
    citecolor={blue!50!black},
    urlcolor={blue!80!black}
}

\usepackage[bitstream-charter]{mathdesign}
\DeclareSymbolFont{usualmathcal}{OMS}{cmsy}{m}{n}
\DeclareSymbolFontAlphabet{\mathcal}{usualmathcal}

\begin{document}

\begin{center}{\Large \textbf{
Optimizing Floquet engineering for non-equilibrium steady states with gradient-based methods
}\\
}\end{center}

\begin{center}
Alberto Castro\textsuperscript{1,2$\star$} and
Shunsuke A. Sato\textsuperscript{3,4}
\end{center}

\begin{center}
{\bf 1} Institute for Biocomputation and Physics of Complex Systems, University of Zaragoza, 50018 Zaragoza (Spain)
\\
{\bf 2} ARAID Foundation, 50018 Zaragoza (Spain)
\\
{\bf 3} Center for Computational Sciences, University of Tsukuba, Tsukuba 305-8577, Japan
\\
{\bf 4} Max Planck Institute for the Structure and Dynamics of Matter, Luruper Chaussee 149, 22761 Hamburg, Germany
\\
${}^\star$ {\small \sf acastro@bifi.es}
\end{center}

\begin{center}
\today
\end{center}


\section*{Abstract}
{\bf
Non-equilibrium steady states are created when a periodically driven
quantum system is also incoherently interacting with an environment --
as it is the case in most realistic situations. The notion of {\it
  Floquet engineering} refers to the manipulation of the properties of
systems under periodic perturbations. Although it more frequently
refers to the coherent states of isolated systems (or to the transient
phase for states that are weakly coupled to the environment), it may
sometimes be of more interest to consider the final steady states that
are reached after decoherence and dissipation take place. In this
work, we demonstrate how those final states can be optimally tuned
with respect to a given predefined metric, such as for example the
maximization of the temporal average value of some observable, by
using multicolor periodic perturbations. We show a computational
framework that can be used for that purpose, and exemplify the concept
using a simple model for the nitrogen-vacancy center in diamond: the
goal in this case is to find the driving periodic magnetic field that
maximizes a time-averaged spin component. We show that, for example,
this technique permits to prepare states whose spin values are
forbidden in thermal equilibrium at any temperature.
}

\vspace{10pt}
\noindent\rule{\textwidth}{1pt}
\tableofcontents\thispagestyle{fancy}
\noindent\rule{\textwidth}{1pt}
\vspace{10pt}

\section{Introduction}
\label{sec:intro}

Exploring novel materials in search of desired properties and functionalities is one of the most important tasks of material sciences and engineering, as it can significantly impact fundamental sciences and practical applications. For example, the conversion efficiency of solar cells has been significantly enhanced over the past several decades through the discovery of various types of materials \cite{doi:10.1126/science.aad4424,Kim2020,https://doi.org/10.1002/aenm.201904102}. Likewise, thanks to the exploration in a vast materials space, various superconducting materials have been found \cite{doi:10.1126/science.288.5465.468,Paglione2010,Cao2018,Drozdov2019}. In addition to these examples, various materials explorations have been conducted toward the realization of desired material properties and functionalities in the equilibrium phase \cite{Varsano2020,Ma2021,Jia2022}.

Recently, the exploration and design of material functionalities has been extended to the nonequilibrium phase of matter under the presence of optical or magnetic drivings. In the seminal work by Oka and Aoki~\cite{PhysRevB.79.081406}, for example, the light-induced anomalous Hall effect in graphene has been theoretically studied in terms of the Floquet picture, suggesting the emergence of topological states of matter. Inspired by this work, various groups have investigated the emergence of new material properties under electromagnetic drivings. The design of material functionalities in the nonequilibrium phase has thus become a full new field of research, that is often called \textit{Floquet engineering}~\cite{PhysRevB.84.235108,Lindner2011,Sentef2015,Hubener2017,doi:10.1146/annurev-conmatphys-031218-013423,PhysRevResearch.4.033213}.

In most theoretical works about Floquet engineering, the states of the target system have been investigated by considering the time-periodic solutions of the Schr\"odinger equation. However, real materials are surrounded by their environment, and those Floquet states, which are the time-periodic solutions of the Schr\"odinger equation, may decay quickly and not be relevant. In fact, recent theoretical and experimental studies suggest that the realization of the Floquet states can be significantly disturbed by their interaction with the environment~\cite{PhysRevE.79.051129,PhysRevB.99.214302,Sato_2020,PhysRevX.10.041013,Aeschlimann2021}. For a practical description of such driven systems, a theory of open-quantum systems under periodic driving has to be considered. 

However, understanding such nonequilibrium phases is significantly
more difficult~\cite{Mori2023}. Nevertheless, a number of works have
addressed the task of computing, characterizing and manipulating
non-equilibrium steady states (NESS). In the realm of condensed matter
systems, Tsuji {\em et al.}~\cite{Tsuji2009}, for example, combined
the Floquet method with nonequilibrium dynamical mean-field theory to
compute NESSs for strongly correlated systems in the presence of both
dissipation and pumping. Dehghani {\em et al.}~\cite{Dehghani2014,
  Dehghani2015} worked on 2D electronic systems irradiated by
cicularly polarized lights and coupled to phonon baths, a problem for
which they developed a kinetic equation approach. Seetharam et
al.~\cite{Seetharam2015} also developed a kinetic equation, which
allowed them to not only compute the NESS, but also to control the
band occupations. The problem of Floquet states occupation control was
also addressed by Iadecola {\em et al}~\cite{Iadecola2015}, who
derived a Floquet master equation for the problem. Murakami {\em et
  al.}~\cite{Murakami2017} used non-equilibrium dynamical mean-field
theory to study the superconducting Holstein model coupled to heat
baths, studying both the transient and the final NESSs.

The possibility of creating novel steady states with the help of
periodic perturbations has also attracted the attention of researchers
in the field of cavity quantum electrodynamics, see for example Shirai
{\em et al.}~\cite{Shirai2014}. Likewise, cold atoms in optical
lattices also form ideal systems for the study of driven
non-equilibrium states and phases, see for example
Refs~\cite{Diehl2008, Diehl2011, Tomita2017, Schnell2023, Wu2022}. We
finish this short sample of previous works that have studied the NESSs
of various systems by stressing the difficulty of computing these
states, even for simple models (see, for example, a recent theoretical
work~\cite{Cheng2022} that has addressed this difficulty, showing how
the use of the matrix product operator approach can help to scale
calculations to larger models).

In this work, we report on the {\it optimization} of the properties of
these NESSs through the design of the time-dependence of the periodic
perturbations. Recently, we have
demonstrated~\cite{PhysRevResearch.4.033213} an approach to Floquet
engineering based on the use of quantum optimal control theory
(QOCT)~\cite{Kirk1998,Shapiro2003,Brif2010,Glaser2015,Castro2018}: the
idea was to allow for multicolor periodic driving, rather than the
monochromatic ones that are normally assumed, and to use the tools of
QOCT to find the amplitudes of the various frequency components that
optimize a given target property of the system -- in that work, the
goal was to modify at will the (pseudo) band structure of graphene.

However, that work also ignored the effect of the environment, and therefore, the found optimal states would only live in a transient prethermalized phase. To realize the Floquet control of material properties and functionalities in systems more tightly coupled to an environment, going beyond the conventional Floquet analysis for isolated systems, we extend here that previous concept of Floquet engineering based on QOCT to open-quantum systems. For this purpose, we first discuss how to apply optimal control theory for nonequilibrium steady states of open-quantum systems under periodic driving, based on a quantum master equation. We then apply the introduced optimal-control procedure to a model of the NV center of diamond under periodic driving, demonstrating that, for example, driven open quantum systems under optimized fields may display exotic properties that are forbidden in the equilibrium phase.

Although to our knowledge, no previous work has attempted the optimization of NESSs with respect to the external drivings, a related work~\cite{VargasHernandez2021} has recently demonstrated the use of automatic differentiation to optimize steady states with respect to internal system or bath parameters.
However, the nature of the problem and the concept of NESS addressed in that work were different
as it was, in consequence, the method. The type of NESSs addressed there are the ones that appear when
a system governed by a static Hamiltonian is in contact with out-of-equilibrium environments (because,
for example, they have different temperatures). The NESS is then time-independent, even if it is not
the one predicted for the quantum canonical ensemble, or any other equilibrium ensemble. 
In our work we are concerned, in contrast,
with the \emph{time-dependent} (and time-periodic) NESSs that appear when a system is driven by some
external fields, while at the same time it is in the presence of a bath. The method described
in Ref.~\cite{VargasHernandez2021} relies on the time-independence of the Lindbladian, and is based
on the application of the implicit function method~\cite{Krantz2003}. Here, we cannot apply that technique,
as one must implicity or explicitly deal with the full propagator in Liouville space.

\section{Method}

In order to manipulate the nonequilibrium steady states, we solve the following optimization problem. Our first assumption is to consider, as master equation, a Lindblad-type equation~\cite{Lindblad1976,Gorini1976} with time-periodic external fields:
\begin{align}
\dot{\rho}(t) = -i\left[H(t), \rho(t)\right] 
+ \sum_{ij} \gamma_{ij} \left( V_{ij}\rho(t)V^\dagger_{ij} - \frac{1}{2}
\lbrace
V_{ij}^\dagger V_{ij}, \rho(t)
\rbrace\right)\,.
\label{eq:lindblad0-eq}
\end{align}
Here, the Hamiltonian $H(t+T) = H(t)$ is periodic with time period $T$. We consider
it to be composed of a field-free and and a periodic perturbation part: $H(t) = H_0 + g(u, t)V$,
where $g(u, t) = g(u, t+T)$ is some $T$-periodic real function parametrized by the
set $u = u_1,\dots, u_P$ -- the {\it control parameters}.
 The incoherent part
of the evolution is determined by the set of Lindblad operators $V_{ij}$, which
we will assume in the following, without loss of generality, 
to be the transition operators $V_{ij} = \vert E_i \rangle\langle E_j\vert$, where
$\vert E_i\rangle$ are the field free Hamiltonian eigenvectors.

We should warn that the previous equation is not universally valid. In
fact, the problem of deriving valid master equations for systems with
time-dependent Hamiltonians is still an open research area~\cite{Mori2023}. The
equation of Lindblad can only be rigorously derived if the Hamiltonian
is time independent -- and even then, it rests on several additional
conditions, most notably Markov's approximation. Various authors have
tackled the problem of deriving master equations for driven
systems~\cite{Albash2012,Davies1978,Lendi1986,Dann2018}.
In some circumstances, Lindblad-type equations with time-dependent
Hamiltonians such as Eq.~(\ref{eq:lindblad0-eq}) are appropriate~\cite{Alicki2007},
and have been used for various purposes~\cite{Ikeda2020,Kienzler2015,Hartmann2017}. 
The previous equation is a simplified version of the so-called Floquet-Lindblad equation~\cite{Ikeda2021}.
We will work with it as working hypothesis; furthermore, the optimization procedured described below can
be easily generalized to more complex master equations.

A Lindblad equation such as the one above can always be written as a linear 
equation in Liouville space:
\begin{equation}
\dot{\rho}(t) = \mathcal{L}(u, t) \rho(t)\,,
\label{eq:lindblad-eq}
\end{equation}
where we now consider $\rho(t)$ to be in vectorized form, i.e it is a $N^2$-dimensional complex vector vector,
where $N$ is the dimension of the underlying Hilbert space~\cite{Havel2003}.
The Lindbladian $\mathcal{L}(u, t)$ is the $N^2\times N^2$ dimensional operator 
that results of transforming Eq.~(\ref{eq:lindblad0-eq}) into
this space. We split it as:
\begin{equation}
\mathcal{L}(u, t) = \mathcal{L}_0 + g(u, t)\mathcal{V}\,.
\end{equation}

Let us call $\rho_u(t)$ to the periodic solution (i.e. $\rho_u(0) =
\rho_u(T)$) of Eq.~(\ref{eq:lindblad-eq}) for a set of parameters
$u$. This solution corresponds to a non-equilibrium steady-state
(NESS). Note that, in principle, there could be more than one steady
state, but we will consider here that it is unique.
We then consider the time-average function
\begin{equation}
F(\rho) = \frac{1}{T}\int_0^T\!{\rm d}t\; \tilde{A}(\rho(t))\,,
\end{equation}
for some function of density matrices $\tilde{A}$ -- in practice, this will typically be the expectation value of some operator $A$:
$\tilde{A}(\rho) = {\rm Tr}[A\rho]$.
The problem that we attempt to solve is the optimization of function:
\begin{equation}
G(u) = F(\rho_u)\,,
\label{eq:opt-target-function}
\end{equation}
subject perhaps to some constraint on the parameters $u$.

Such class of optimization problems for time-dependent processes that
can be controlled by the manipulation of external handles is the
object of (quantum, in this case) optimal control theory (QOCT). Any
function optimization algorithm requires a 
method for the computation of the function;
in addition, many efficient
algorithms will also require a method for the computation of its gradient. Computing the function $G$ essentially amounts
to obtaining the NESS. In the
following, we will show one possible way to do this, and also derive one expression for the gradient. Note that
since
\begin{equation}
\label{eq:functionG}
G(u) = \frac{1}{T}\int_0^T\!\!{\rm d}t\; {\rm Tr}[A \rho_u(t)],
\end{equation}
the gradient components may then be computed as:
\begin{equation}
\frac{\partial G}{\partial u_k} = \frac{1}{T}\int_0^T\!\!{\rm d}t\;
{\rm Tr}[A \frac{\partial \rho_u}{\partial u_k}(t)],
\label{eq:gradient-G}
\end{equation}
and therefore the problem in fact amounts to finding some procedure to compute
the derivatives $\frac{\partial \rho_u}{\partial u_k}$.

Let us first rewrite Eq.~(\ref{eq:lindblad-eq}) elementwise:
\begin{equation}
\dot{\rho}_\alpha(t) = \sum_\beta \mathcal{L}_{\alpha\beta}(u, t)\rho_\beta(t)
\end{equation}
and consider the Fourier transform of these objects:
\begin{eqnarray}
\rho_\alpha(t) &=& \sum_{n} \rho_{\alpha, n}e^{i\omega_n t},
\\
\rho_{\alpha, n} &=& \frac{1}{T}\int_0^T\!\!{\rm d}t\; e^{-i\omega_n t}\rho_{\alpha}(t),
\\
\mathcal{L}_{\alpha\beta}(u, t) &=& \sum_{n} \mathcal{L}_{\alpha\beta, n}(u)e^{i\omega_n t},
\\
\mathcal{L}_{\alpha\beta, n}(u) &=& \frac{1}{T}\int_0^T\!\!{\rm d}t\; e^{-i\omega_n t}\mathcal{L}_{\alpha\beta}(u, t),
\end{eqnarray}
where $\omega_n = \frac{2\pi}{T}n\,,\quad n=0,1,\dots,N-1$. In the frequency domain,
the Lindblad equation, Eq.~(\ref{eq:lindblad-eq}), can then be rewritten as~\footnote{
These equations are easily reached using the following two formulas:
$$
\frac{1}{T}\int_0^T\!\!{\rm d}t\; \dot{\rho}_\alpha(t)e^{-i\omega_n t} = i\omega_n \rho_{\alpha, n},
$$
and
$$
\frac{1}{T}\int_0^T\!\!{\rm d}t\; \mathcal{L}_{\alpha\beta}(u, t)\rho_\beta(t)e^{-i\omega_n t} = 
\sum_{n=0}^{N-1} \mathcal{L}_{\alpha\beta, n-m}(u)\rho_{\beta,m}.
$$
}:
\begin{equation}
\sum_\beta\sum_{m=0}^{N-1}\left[ \mathcal{L}_{\alpha\beta,n-m}(u)-i\delta_{nm}\delta_{\alpha\beta}\omega_m\right]\rho_{\beta,m} = 0.
\end{equation}
And, by further defining the following operator
\begin{equation}
\overline{\mathcal{L}}_{\alpha n,\beta m}(u) = \mathcal{L}_{\alpha\beta,n-m}(u) -i\delta_{nm}\delta_{\alpha\beta}\omega_m,
\end{equation}
we finally rewrite Eq.~(\ref{eq:lindblad-eq}) as:
\begin{equation}
\label{eq:homogeneous}
\sum_\beta \sum_{m=0}^{N-1} \overline{\mathcal{L}}_{\alpha n,\beta m}(u) \rho_{\beta, m} = 0.
\end{equation}

This is a linear homogeneous equation; the solution
(the nullspace or kernel, assuming that it has dimension one), will be
the periodic solution that we are after, the NESS~\footnote{
Other procedures could be used to compute the NESS, sometimes also called
``asymptotic Floquet states'', such as for example simply propagating the equation for a long
time, as the system should decay to the steady state.
}. 
We now need some procedure to find $\frac{\partial \rho}{\partial u_k}$. Taking variations of Eq.~(\ref{eq:homogeneous}) with respect to the parameters $u$, we get:
\begin{equation}
\label{eq:mainequation}
\overline{\mathcal{L}}(u)\frac{\partial \rho}{\partial u_k}(u) = - \frac{\partial \overline{\mathcal{L}}}{\partial u_m}(u)\rho_u.
\end{equation}
This is a linear equation that would provide $\frac{\partial
  \rho_u}{\partial u_k}$. However, note that since $\overline{\mathcal{L}}(u)$ has
a non-empty kernel (given precisely by $\rho_u$), it cannot be solved
straightforwardly. In fact, it does not have a unique solution: If $x$
is a solution of
\begin{equation}
\overline{\mathcal{L}}(u)x = - \frac{\partial \overline{\mathcal{L}}}{\partial u_m}(u)\rho(u),
\label{eq:compute-x}
\end{equation}
$x + \mu \rho_u$ is also a solution for any $\mu$.
To remove this arbitrariness, we impose the normalization condition, ${\rm Tr}\rho_u = 1$ for any $u$, and therefore:
\begin{equation}
{\rm Tr} \frac{\partial \rho_u}{\partial u_k} = 0.
\label{eq:condition-for-gradient}
\end{equation}
To find $\frac{\partial \rho_u}{\partial u_k}$ in practice, we may then take the following two steps: First, we compute a solution of the linear equation, Eq.~(\ref{eq:compute-x}), with the least-squares method, by imposing that the solution $x_0$ is perpendicular to the kernel, i.e.: $x_0^\dagger \cdot \rho_u = 0$. Then, we update the solution with the condition, Eq.~(\ref{eq:condition-for-gradient}). The required solution is obtained as:
\begin{equation}
\frac{\partial \rho_u}{\partial u_k} = x_0 - ({\rm Tr} x_0) \rho_u.
\end{equation}
Once we have $\frac{\partial \rho_u}{\partial u_k}$, we can evaluate the gradient 
in Eq.~(\ref{eq:gradient-G}). Armed with this procedure to compute this gradient,
one can perform the optimization of function $G(u)$ with many efficient algorithms.
This method has been implemented in the qocttools code~\cite{qocttools}, publicly
available, and all the necessary scripts and data necessary to replicate the following results
are also available upon request from the authors. 

While Eq.~(\ref{eq:mainequation}) may seem similar to Eqs. (4-5) in Ref.~\cite{VargasHernandez2021},
it is not the same equation: The linear operator here is defined in Floquet-Liouville space, as it is an extended
Lindbladian to that space, due to the time-dependence of the problem. The linear equations (4-5) in
Ref.~\cite{VargasHernandez2021} (see also \cite{VargasHernandez2020}) 
are derived from the assumption of a time-independent NESS:
\begin{equation}
\mathcal{L}\rho = 0\,,
\end{equation}
an assumption that we cannot make in our context.

We note that this procedure to obtain the gradient is not the typical
route followed when working with QOCT, that is normally based on
equations deduced from Pontryagin's maximum
principle~\cite{Boltyanskii1956, Pontryagin1962, PRXQuantum.2.030203}
or, relatedly, on the adjoint method~\cite{Cao2003}. While we have
attempted the use of those two methods, we did not obtain satisfactory
results. The reason is the difficulty in finding a solution to the
adjoint equations. Both approaches rely on the definion of an adjoint
state (also known as \emph{costate}), defined by an auxiliary
equation. In the case of periodic equations, the solution to that
adjoint equation must be periodic, too. While finding solutions to
initial value problems, required for normal QOCT problems, is often
easy and their existence is guaranteed, this is not the case for
periodic equations. We have found difficulties in finding solutions to
the adjoint equation, and opted for developing the technique described
here.

Both the calculation of the NESS, and of its gradient (which are the
ingredients of the optimization process) rely on obtaining solutions
to algebraic linear systems of dimension $M = d^2N$, where $d$ is the
Hilbert space dimension, and $N$ is the dimension of the \emph{time
space} (the number of frequencies in the Fourier expansion, or the
number of time steps in the time discretization). These linear systems
must be solved at least once for each iteration step in any
optimization algorithm, at a cost that, in the general case, grows
as $M^3$. This can imply a significant computational cost. In order to scale the
method to large systems, one should make use of sparse representations
of the Hamiltonian (which will become sparse Liouvillians). In this
way, the complexity of the problem is greatly reduced $M^2$. Even
then, we do not expect that the current method can be easily applied
directly on top of first principles techniques used for many-body
interacting systems. Instead, the route would consist in
constructing a model out of the first principles calculations,
performing the optimization with the model, and testing the output
back with the first principles method.

\section{Results}

In the following, we will use the previous equations with 
the following model of the NV center of diamond~\cite{Ikeda2020,Rondin2014}:
\begin{eqnarray}
H(u, t) &=& H_0 + V(u, t),
\\
H_0 &=& -B_s S_z + N_z S_z^2 + N_{xy}(S_x^2-S_y^2),
\\
\label{eq:tdpart}
V(u, t) &=& - g_x(t) B_d S_x - g_y(t) B_d S_y.
\end{eqnarray}
The model definition must be completed with the definition of the dissipative part: we take $\gamma_{ij} = \gamma e^{-\beta E_i} / (e^{-\beta E_i}+e^{-\beta E_j})$ and $\gamma_{ii}=0$, where $\beta = 1/(k_{\rm B}T)$ is the inverse of the temperature, and $\gamma$ is a rate constant~\footnote{Notice that this dissipation model ensures the detailed balance condition, $\gamma_{ij}e^{-\beta E_j} = \gamma_{ji}e^{-\beta E_i}$.}. The reason for choosing this model is the work of Ikeda {\it et al.}~\cite{Ikeda2020}, who studied the NESSs of this system under circularly polarized light ($g_x(t) = \cos(\omega t)$; $g_y(t) = \sin(\omega t)$). In that work, the high-frequency approximation was used in order to derive simplified expressions for the NESS. Here, the goal would be to parametrize functions $g_x = g_x(u, t)$ and $g_y = g_y(u, t)$, and find the parameters $u$ that result in a NESS that maximizes the time-averaged value of some observable (for example, $S_z$).

Following Ikeda {\em et al.}~\cite{Ikeda2020}, we set the units of the model by fixing $N_z = 1$; the rest of the parameters of the model are then given by: $N_{\rm xy} = 0.05, B_s = 0.3, B_d = 0.1, \gamma = 0.2$ (see \cite{Rondin2014} for a review on the NV diamond centers, this and other models, and the typical values that these constants may take).

\begin{figure}
\centerline{\includegraphics[width=0.8\columnwidth]{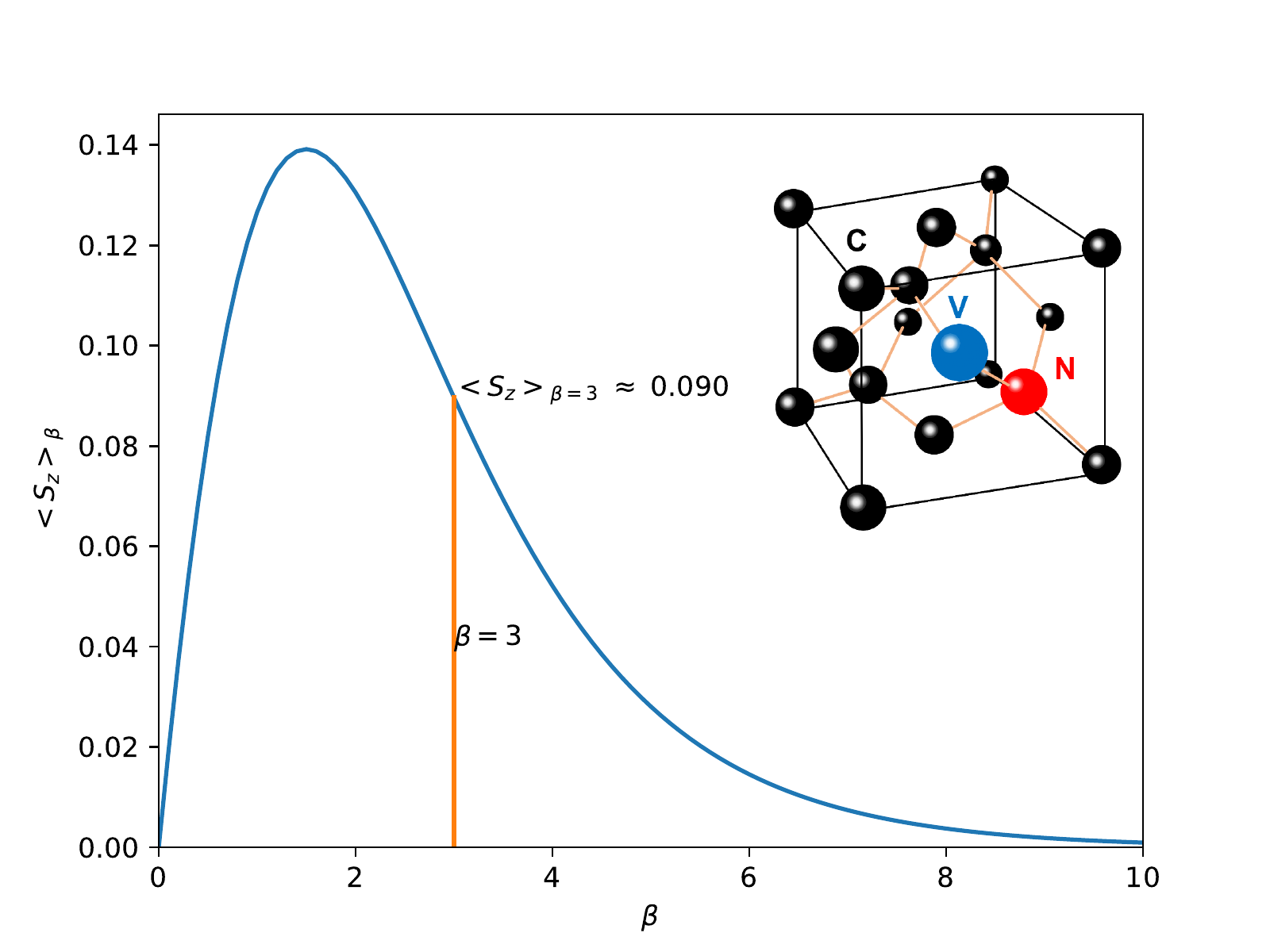}}
\caption{\label{fig:ovsbeta}
Thermal average of $S_z$, as a function of the inverse temperature $\beta = \frac{1}{k_{\rm B} T}$. The value at $\beta = 3$, used in the text for the rest of the calculations, is singled out. {\it Inset}: structure of the Nitrogen vacancy defect in diamond.
}
\end{figure}
First, let us consider the field-free value of $S_z$; the thermal average of $S_z$, $\langle S_z\rangle_\beta$, is shown in Fig.~\ref{fig:ovsbeta} as a function of the inverse temperature $\beta$. One can see how at zero temperature $(\beta \to \infty)$, $\langle S_z\rangle_\beta \to 0$, reflecting the fact that the ground-state value of $S_z$ is also zero: $\langle \psi_0 \vert S_z\vert\psi_0\rangle = 0$. As the temperature increases, the population of the first excited state grows, and therefore the thermal average of $S_z$ also grows, since $\langle \psi_1 \vert S_z \vert \psi_1\rangle \approx 1$. However, if the temperature is increased further, the population of the second excited state also starts to grow, and the thermal average starts to decrease, as $\langle\psi_2 \vert S_z \vert \psi_2\rangle \approx -1$. In the limit of infinite temperature ($\beta \to 0$), the thermal average approaches zero again, as that limit involves an equally populated ensemble of all three states. Note then that a {\it thermal control} of $S_z$, i.e. the manipulation of the value of $S_z$ via a variation of the temperature, is limited to the range $0 < \langle S_z\rangle_\beta < 0.14$.

However, as we will show, if a periodic perturbation is added, this range can be enlarged, and one may reach NESSs with larger or smaller values of the (time averaged) $S_z$. In the following, let us fix $\beta = 3$, and seek for the drivings that are capable of producing those NESSs.
The first step is to set a parametrized form for the time-dependent functions $g_x$ and $g_y$ used in Eq.~(\ref{eq:tdpart}); the simplest choice is to use Fourier expansions:
\begin{eqnarray}
g_x(u, t) &=& u_0 + \sum_{n=1}^M \left[ u_{2n}\cos(\omega_n t) + u_{2n-1}\sin(\omega_n t)\right]\,,
\\\nonumber
g_y(u, t) &=& u_{2M+1} + \sum_{n=1}^M \left[ u_{2M+1+2n}\cos(\omega_n t) + u_{2M+2n}\sin(\omega_n t)\right]\,.
\end{eqnarray}

The control parameters are therefore the Fourier coefficients of the temporal shape of the two magnetic fields, $u_0,\dots u_{4M+1}$. The index $M$ determines the cutoff frequency $\omega_M$, whereas all the Fourier frequencies are $\omega_n = n\omega_0$ for $n=1,\dots,M$. A choice must then be made on the fundamental frequency $\omega_0$, which is of course related to the period that we choose for the external field $\omega_0 = \frac{2\pi}{T}$. In this work, we have chosen $\omega_0 = 0.5~N_z$, and $M=4$, such that the cutoff frequency is $\omega_M = 2.0~N_z$. By defining the control functions in this parametrized manner, we effectively constrain the final solution to a given domain of validity -- in this case setting a maximum frequency. This would be consistent with any experimental realization of this concept, as in practice the time-dependent magnetic fields would also be constrained in frequencies due to technological limitations.

The optimization of function (\ref{eq:functionG}) may then be started using any gradient-based algorithm -- the one that we have used for these calculations is the Sequential Least-Squares Quadratic Programming (SLSQP) algorithm~\cite{Kraft1994} as implemented in the NLOPT library~\cite{nlopt}. Note that we have not performed an unconstrained maximization for all possible values of parameters $u_j$, but we have added a constraint on the amplitudes of each frequency component:
\begin{equation}
\vert u_j \vert \le \kappa\quad\textrm{for any }j.
\end{equation}
Such a constraint would also be present in an experiment. The chosen algorithm permits to include this constraint
-- this was our reason for choosing it; however note that many other algorithms also allow for
bounds and linear or nonlinear constraints. We did not attempt any study of the relative performances of
different algorithms for this particular type of problem, leaving that study for a future work.

\begin{figure}
\centerline{\includegraphics[width=0.49\columnwidth]{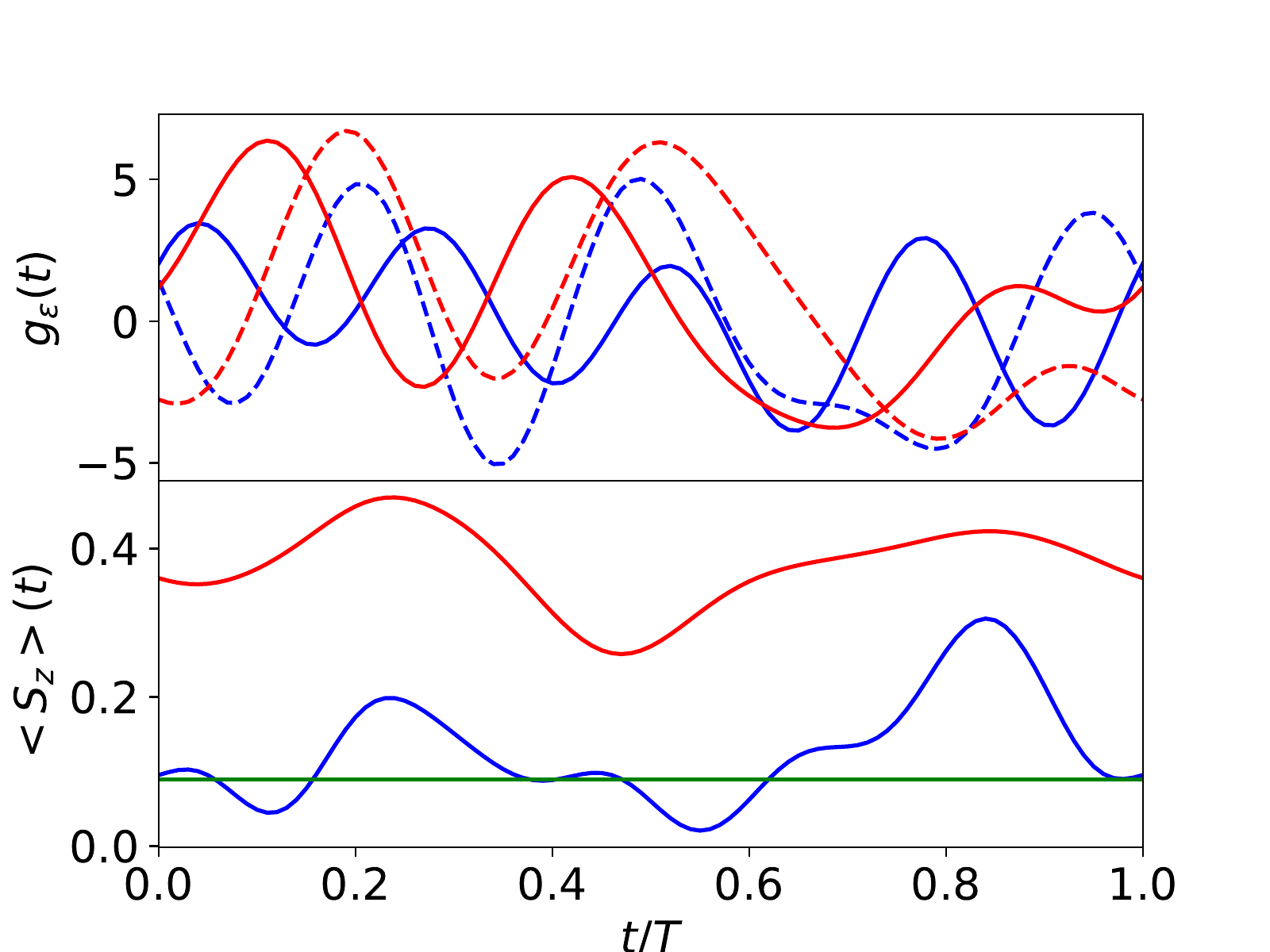}
            \includegraphics[width=0.49\columnwidth]{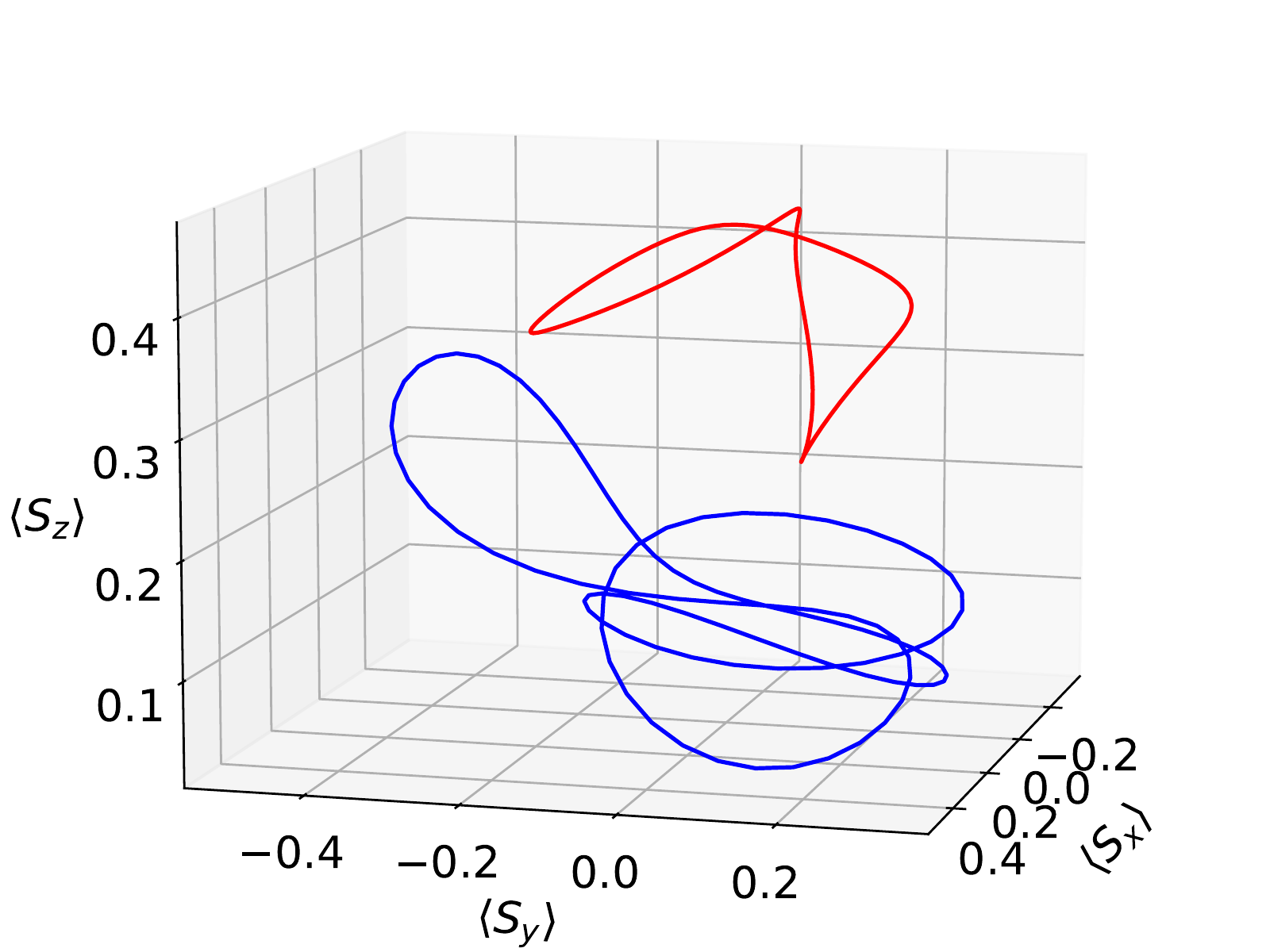}}
\caption{\label{fig:refoptimization}
{\it Left, top:} Optimized (red) and initial guess (blue) temporal shapes of the time-dependent magnetic fields $g_x$ (solid) and $g_y$ (dashed). {\it Left, bottom:} Evolution of $\langle S_z\rangle$ when using the initial guess (blue) and the optimal fields (red). The green line represent the thermal average at $\beta = 3$. {\it Right:} Trajectories of the spin vector $\langle \vec{S}(t)\rangle$ during one period $T$, for the initial guess (blue) and optimized (red) perturbations.
}
\end{figure}

Fig.~\ref{fig:refoptimization} shows the results of one optimization; in this case the amplitudes were constrained using $\kappa = 4.0$.  The optimization is started with random fields (shown in the left, top panel, with blue lines), and then proceeds iteratively until the fields that optimize the temporal average of $S_z$ are found (shown in the left, top panel, with red lines). In the left, bottom panel, the evolutions in time of $S_z$ are shown, once again for the initial guess and for the optimized case. It can be seen how the optimized fields lead to significantly higher values of $S_z$ -- both with respect to the initial random fields, and with respect to the thermal value (shown as a straight green line in the plot). In fact, the time-averaged value of $S_z$ achieved in this way ($\approx 0.38$) is higher than the maximum that can be achieved in equilibrium phase by modifyng the temperature ($\approx 0.14$, as discussed above). The right part of Fig.~\ref{fig:refoptimization} shows the full spin vector $\langle \vec{S}(t)\rangle$ evolving in time during one Floquet period, both for the initial (blue) and optimized (red) cases.

\begin{figure}
\centerline{\includegraphics[width=0.49\columnwidth]{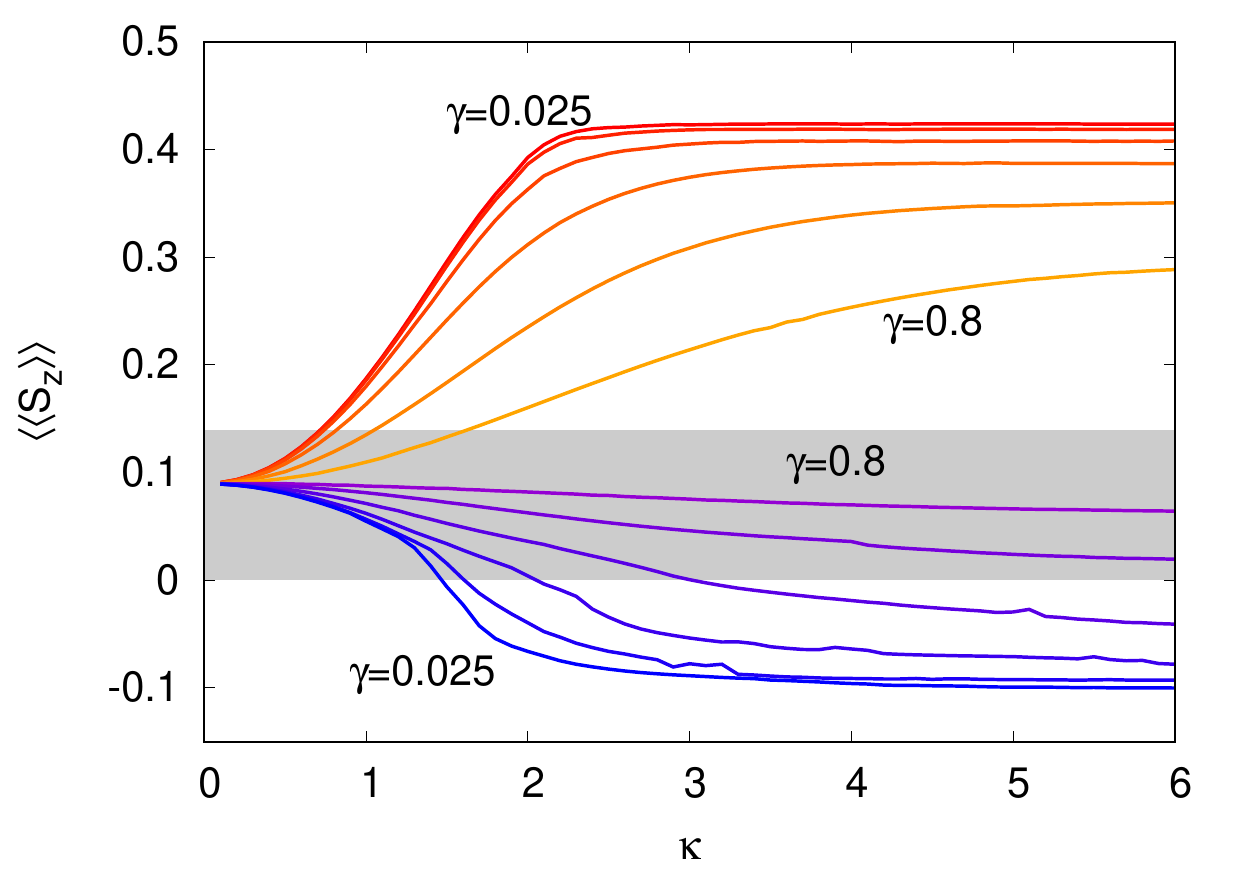}
            \includegraphics[width=0.49\columnwidth]{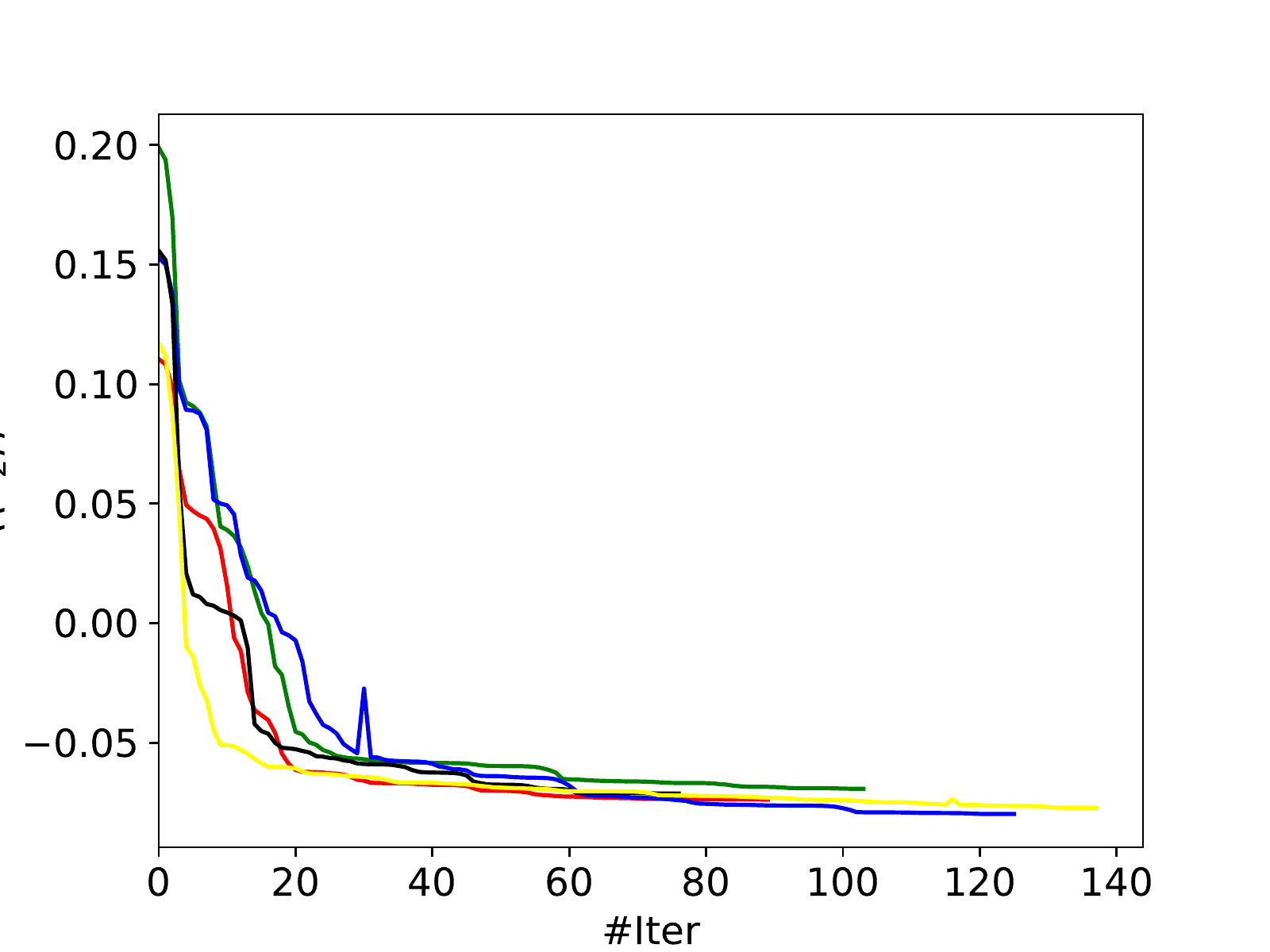}}
\caption{\label{fig:szvsbound}
{\it Left:} Maximized (red-orange) and minimized (blue-violet) values of the time-averaged $S_z$ expectation value, $\langle\langle S_z\rangle\rangle$, as a function of the amplitude bound $\kappa$. The various curves correspond to different values of the rate constant $\gamma$, which are doubled from $\gamma = 0.025$ to $\gamma = 0.8$. The shaded region marks the only allowed values of $S_z$ in thermal equilibrium (thus, for example $\langle S_z\rangle_\beta > 0$).
{\it Right:} Convergence histories for some of the optimization runs: five optimizations with different random
initial guesses; for all of them, the target was the minimization of $\langle\langle S_z\rangle\rangle$,
for $\gamma = 0.025, \kappa = 2$.
}
\end{figure}

The final optimized value of function $G$ (i.e. of the time averaged value of $S_z$) obviously depends on how we constrain the periodic functions. For example, on the bound $\kappa$ that we set on the amplitudes. Fig.~\ref{fig:szvsbound} shows the optimal value obtained as a function of that bound (red curves), for various values of the dissipation constant $\gamma$. Obviously, if the bound is set to a very small value, the presence of the periodic field barely modifies the thermal average (of around 0.09, for the chosen temperature value, $\beta = 3$). However, if the bound is relaxed to higher values, the average can be significantly increased, up to a saturation value that depends on $\gamma$: the higher the $\gamma$, the lower the value of the optimized $\langle\langle S_z\rangle\rangle$. This can be understood physically, as a faster dissipation drives with more strength the system towards its thermal equilibrium state. Finally, we have attempted to {\it minimize} the time average of $S_z$, wondering whether one can engineer states with the in principle forbidden negative spin values. In Fig.~\ref{fig:szvsbound} we display the obtained optimal values, also as a function of the amplitude bound (red curves). It may be seen how, if sufficiently big amplitudes are allowed, one may actually obtain negative values -- which are forbidden in thermal equilibrium, as it can be seen in Fig.~\ref{fig:ovsbeta}.

Finally, a word about the optimization process: for each case, we did several runs initiated from
different randomly generated initial guess pulses. This procedure demonstrated that the optimization problem
has in fact multiple local maxima; the plotted data points in Fig.~\ref{fig:szvsbound} (left) are the best results
found in each case. The optimization algorithm itself requires around 100 evaluations of the gradient and of
the function to reach a well converged result; Fig.~\ref{fig:szvsbound} (right) shows some typical convergence
histories for several different random initial guesses -- in this case, for the 
$\gamma = 0.025, \kappa = 2$ case (the target was the minimization of
$\langle\langle S_z\rangle\rangle$).

\section{Conclusions}

We have developed an optimal control scheme for the nonequilibrium steady states of open quantum systems under time-periodic drivings, aiming to control the properties of matter in nonequilibrium phases. We derived an expression for the gradient vectors of physical observables in NESSs with respect to the parameters of the external periodic fields, and we employed these derived gradient vectors for the optimization of observables of the diamond NV center under external periodic magnetic fields. We confirmed that the time-averaged value of the spin component, $S_z$, can be controled with the proposed optimal control sheme. Furthermore, we demonstrated that this technique can be used to find ``exotic'' NESSs, such as states that display properties that are forbidden in equilibrium phases: As shown in Fig.~\ref{fig:szvsbound}, the $z$-spin component of the optimized NESS can be outside the range of values allowed in equilibrium -- for example, it may be negative, which is impossible at any temperature.

Having established an optimal control scheme for NESSs under periodic driving, the field parameters can be added as novel degrees of freedom for material explorations aimed to endow the materials with desired properties and functionalities. This extends the concept of material exploration, from equilibrium to nonequilibrium situations. Because the present optimization scheme is based on the steady state solutions of a master equation, such as Lindblad's equation [Eq.~(\ref{eq:lindblad0-eq})], the relaxation and dissipation effects are naturally included in the optimization procedure. Hence, the engineering of material properties based on the proposed scheme can be seen as an extension of the more common Floquet engineering usually based on the steady solutions of the time-dependent Schr\"odinger equation without taking into account the relaxation and dissipation effects. The optimal control of NESSs proposed in this work shows how the difficulties of Floquet engineering due to the relaxation and dissipation effects can be overcome, and the natural inclusion of these effects opens a path to the control of material properties with experimentally realizable fields.

\section*{Acknowledgements}


\paragraph{Funding information}
AC acknowledges support from Grant PID2021-123251NB-I00 funded by MCIN/AEI/10.13039/501100011033. SAS acknowledges the support from JSPS KAKENHI Grant Numbers JP20K14382.

\bibliography{refs.bib}

\begin{thebibliography}{10}
\providecommand{\url}[1]{\texttt{#1}}
\providecommand{\urlprefix}{URL }
\expandafter\ifx\csname urlstyle\endcsname\relax
  \providecommand{\doi}[1]{doi:\discretionary{}{}{}#1}\else
  \providecommand{\doi}{doi:\discretionary{}{}{}\begingroup
  \urlstyle{rm}\Url}\fi
\providecommand{\eprint}[2][]{\url{#2}}

\bibitem{doi:10.1126/science.aad4424}
A.~Polman, M.~Knight, E.~C. Garnett, B.~Ehrler and W.~C. Sinke,
\newblock \emph{Photovoltaic materials: Present efficiencies and future
  challenges},
\newblock Science \textbf{352}(6283), aad4424 (2016),
\newblock \doi{10.1126/science.aad4424}.

\bibitem{Kim2020}
J.~Y. Kim, J.-W. Lee, H.~S. Jung, H.~Shin and N.-G. Park,
\newblock \emph{High-efficiency perovskite solar cells},
\newblock Chemical Reviews \textbf{120}(15), 7867 (2020),
\newblock \doi{10.1021/acs.chemrev.0c00107}.

\bibitem{https://doi.org/10.1002/aenm.201904102}
M.~Jošt, L.~Kegelmann, L.~Korte and S.~Albrecht,
\newblock \emph{Monolithic perovskite tandem solar cells: A review of the
  present status and advanced characterization methods toward 30\% efficiency},
\newblock Advanced Energy Materials \textbf{10}(26), 1904102 (2020),
\newblock \doi{https://doi.org/10.1002/aenm.201904102}.

\bibitem{doi:10.1126/science.288.5465.468}
J.~Orenstein and A.~J. Millis,
\newblock \emph{Advances in the physics of high-temperature superconductivity},
\newblock Science \textbf{288}(5465), 468 (2000),
\newblock \doi{10.1126/science.288.5465.468}.

\bibitem{Paglione2010}
J.~Paglione and R.~L. Greene,
\newblock \emph{High-temperature superconductivity in iron-based materials},
\newblock Nature Physics \textbf{6}(9), 645 (2010),
\newblock \doi{10.1038/nphys1759}.

\bibitem{Cao2018}
Y.~Cao, V.~Fatemi, S.~Fang, K.~Watanabe, T.~Taniguchi, E.~Kaxiras and
  P.~Jarillo-Herrero,
\newblock \emph{Unconventional superconductivity in magic-angle graphene
  superlattices},
\newblock Nature \textbf{556}(7699), 43 (2018),
\newblock \doi{10.1038/nature26160}.

\bibitem{Drozdov2019}
A.~P. Drozdov, P.~P. Kong, V.~S. Minkov, S.~P. Besedin, M.~A. Kuzovnikov,
  S.~Mozaffari, L.~Balicas, F.~F. Balakirev, D.~E. Graf, V.~B. Prakapenka,
  E.~Greenberg, D.~A. Knyazev \emph{et~al.},
\newblock \emph{Superconductivity at 250 k in lanthanum hydride under high
  pressures},
\newblock Nature \textbf{569}(7757), 528 (2019),
\newblock \doi{10.1038/s41586-019-1201-8}.

\bibitem{Varsano2020}
D.~Varsano, M.~Palummo, E.~Molinari and M.~Rontani,
\newblock \emph{A monolayer transition-metal dichalcogenide as a topological
  excitonic insulator},
\newblock Nature Nanotechnology \textbf{15}(5), 367 (2020),
\newblock \doi{10.1038/s41565-020-0650-4}.

\bibitem{Ma2021}
L.~Ma, P.~X. Nguyen, Z.~Wang, Y.~Zeng, K.~Watanabe, T.~Taniguchi, A.~H.
  MacDonald, K.~F. Mak and J.~Shan,
\newblock \emph{Strongly correlated excitonic insulator in atomic double
  layers},
\newblock Nature \textbf{598}(7882), 585 (2021),
\newblock \doi{10.1038/s41586-021-03947-9}.

\bibitem{Jia2022}
Y.~Jia, P.~Wang, C.-L. Chiu, Z.~Song, G.~Yu, B.~J{\"a}ck, S.~Lei, S.~Klemenz,
  F.~A. Cevallos, M.~Onyszczak, N.~Fishchenko, X.~Liu \emph{et~al.},
\newblock \emph{Evidence for a monolayer excitonic insulator},
\newblock Nature Physics \textbf{18}(1), 87 (2022),
\newblock \doi{10.1038/s41567-021-01422-w}.

\bibitem{PhysRevB.79.081406}
T.~Oka and H.~Aoki,
\newblock \emph{Photovoltaic hall effect in graphene},
\newblock Phys. Rev. B \textbf{79}, 081406 (2009),
\newblock \doi{10.1103/PhysRevB.79.081406}.

\bibitem{PhysRevB.84.235108}
T.~Kitagawa, T.~Oka, A.~Brataas, L.~Fu and E.~Demler,
\newblock \emph{Transport properties of nonequilibrium systems under the
  application of light: Photoinduced quantum hall insulators without landau
  levels},
\newblock Phys. Rev. B \textbf{84}, 235108 (2011),
\newblock \doi{10.1103/PhysRevB.84.235108}.

\bibitem{Lindner2011}
N.~H. Lindner, G.~Refael and V.~Galitski,
\newblock \emph{Floquet topological insulator in semiconductor quantum wells},
\newblock Nature Physics \textbf{7}(6), 490 (2011),
\newblock \doi{10.1038/nphys1926}.

\bibitem{Sentef2015}
M.~A. Sentef, M.~Claassen, A.~F. Kemper, B.~Moritz, T.~Oka, J.~K. Freericks and
  T.~P. Devereaux,
\newblock \emph{Theory of floquet band formation and local pseudospin textures
  in pump-probe photoemission of graphene},
\newblock Nature Communications \textbf{6}(1), 7047 (2015),
\newblock \doi{10.1038/ncomms8047}.

\bibitem{Hubener2017}
H.~H{\"u}bener, M.~A. Sentef, U.~De~Giovannini, A.~F. Kemper and A.~Rubio,
\newblock \emph{Creating stable floquet--weyl semimetals by laser-driving of 3d
  dirac materials},
\newblock Nature Communications \textbf{8}(1), 13940 (2017),
\newblock \doi{10.1038/ncomms13940}.

\bibitem{doi:10.1146/annurev-conmatphys-031218-013423}
T.~Oka and S.~Kitamura,
\newblock \emph{Floquet engineering of quantum materials},
\newblock Annual Review of Condensed Matter Physics \textbf{10}(1), 387 (2019),
\newblock \doi{10.1146/annurev-conmatphys-031218-013423}.

\bibitem{PhysRevResearch.4.033213}
A.~Castro, U.~De~Giovannini, S.~A. Sato, H.~H\"ubener and A.~Rubio,
\newblock \emph{Floquet engineering the band structure of materials with
  optimal control theory},
\newblock Phys. Rev. Research \textbf{4}, 033213 (2022),
\newblock \doi{10.1103/PhysRevResearch.4.033213}.

\bibitem{PhysRevE.79.051129}
D.~W. Hone, R.~Ketzmerick and W.~Kohn,
\newblock \emph{Statistical mechanics of floquet systems: The pervasive problem
  of near degeneracies},
\newblock Phys. Rev. E \textbf{79}, 051129 (2009),
\newblock \doi{10.1103/PhysRevE.79.051129}.

\bibitem{PhysRevB.99.214302}
S.~A. Sato, J.~W. McIver, M.~Nuske, P.~Tang, G.~Jotzu, B.~Schulte,
  H.~H\"ubener, U.~De~Giovannini, L.~Mathey, M.~A. Sentef, A.~Cavalleri and
  A.~Rubio,
\newblock \emph{Microscopic theory for the light-induced anomalous hall effect
  in graphene},
\newblock Phys. Rev. B \textbf{99}, 214302 (2019),
\newblock \doi{10.1103/PhysRevB.99.214302}.

\bibitem{Sato_2020}
S.~A. Sato, U.~D. Giovannini, S.~Aeschlimann, I.~Gierz, H.~Hübener and
  A.~Rubio,
\newblock \emph{Floquet states in dissipative open quantum systems},
\newblock Journal of Physics B: Atomic, Molecular and Optical Physics
  \textbf{53}(22), 225601 (2020),
\newblock \doi{10.1088/1361-6455/abb127}.

\bibitem{PhysRevX.10.041013}
M.~Sch\"uler, U.~De~Giovannini, H.~H\"ubener, A.~Rubio, M.~A. Sentef, T.~P.
  Devereaux and P.~Werner,
\newblock \emph{How circular dichroism in time- and angle-resolved
  photoemission can be used to spectroscopically detect transient topological
  states in graphene},
\newblock Phys. Rev. X \textbf{10}, 041013 (2020),
\newblock \doi{10.1103/PhysRevX.10.041013}.

\bibitem{Aeschlimann2021}
S.~Aeschlimann, S.~A. Sato, R.~Krause, M.~Ch{\'a}vez-Cervantes,
  U.~De~Giovannini, H.~H{\"u}bener, S.~Forti, C.~Coletti, K.~Hanff,
  K.~Rossnagel, A.~Rubio and I.~Gierz,
\newblock \emph{Survival of floquet--bloch states in the presence of
  scattering},
\newblock Nano Letters \textbf{21}(12), 5028 (2021),
\newblock \doi{10.1021/acs.nanolett.1c00801}.

\bibitem{Mori2023}
T.~Mori,
\newblock \emph{Floquet states in open quantum systems},
\newblock Annual Review of Condensed Matter Physics \textbf{14}(1), 35 (2023),
\newblock \doi{10.1146/annurev-conmatphys-040721-015537}.

\bibitem{Tsuji2009}
N.~Tsuji, T.~Oka and H.~Aoki,
\newblock \emph{Nonequilibrium steady state of photoexcited correlated
  electrons in the presence of dissipation},
\newblock Phys. Rev. Lett. \textbf{103}, 047403 (2009),
\newblock \doi{10.1103/PhysRevLett.103.047403}.

\bibitem{Dehghani2014}
H.~Dehghani, T.~Oka and A.~Mitra,
\newblock \emph{Dissipative floquet topological systems},
\newblock Phys. Rev. B \textbf{90}, 195429 (2014),
\newblock \doi{10.1103/PhysRevB.90.195429}.

\bibitem{Dehghani2015}
H.~Dehghani, T.~Oka and A.~Mitra,
\newblock \emph{Out-of-equilibrium electrons and the hall conductance of a
  floquet topological insulator},
\newblock Phys. Rev. B \textbf{91}, 155422 (2015),
\newblock \doi{10.1103/PhysRevB.91.155422}.

\bibitem{Seetharam2015}
K.~I. Seetharam, C.-E. Bardyn, N.~H. Lindner, M.~S. Rudner and G.~Refael,
\newblock \emph{Controlled population of floquet-bloch states via coupling to
  bose and fermi baths},
\newblock Phys. Rev. X \textbf{5}, 041050 (2015),
\newblock \doi{10.1103/PhysRevX.5.041050}.

\bibitem{Iadecola2015}
T.~Iadecola, T.~Neupert and C.~Chamon,
\newblock \emph{Occupation of topological floquet bands in open systems},
\newblock Phys. Rev. B \textbf{91}, 235133 (2015),
\newblock \doi{10.1103/PhysRevB.91.235133}.

\bibitem{Murakami2017}
Y.~Murakami, N.~Tsuji, M.~Eckstein and P.~Werner,
\newblock \emph{Nonequilibrium steady states and transient dynamics of
  conventional superconductors under phonon driving},
\newblock Phys. Rev. B \textbf{96}, 045125 (2017),
\newblock \doi{10.1103/PhysRevB.96.045125}.

\bibitem{Shirai2014}
T.~Shirai, T.~Mori and S.~Miyashita,
\newblock \emph{Novel symmetry-broken phase in a driven cavity system in the
  thermodynamic limit},
\newblock Journal of Physics B: Atomic, Molecular and Optical Physics
  \textbf{47}(2), 025501 (2013),
\newblock \doi{10.1088/0953-4075/47/2/025501}.

\bibitem{Diehl2008}
S.~Diehl, A.~Micheli, A.~Kantian, B.~Kraus, H.~P. B{\"u}chler and P.~Zoller,
\newblock \emph{Quantum states and phases in driven open quantum systems with
  cold atoms},
\newblock Nature Physics \textbf{4}(11), 878 (2008),
\newblock \doi{10.1038/nphys1073}.

\bibitem{Diehl2011}
S.~Diehl, E.~Rico, M.~A. Baranov and P.~Zoller,
\newblock \emph{Topology by dissipation in atomic quantum wires},
\newblock Nature Physics \textbf{7}(12), 971 (2011),
\newblock \doi{10.1038/nphys2106}.

\bibitem{Tomita2017}
T.~Tomita, S.~Nakajima, I.~Danshita, Y.~Takasu and Y.~Takahashi,
\newblock \emph{Observation of the mott insulator to superfluid crossover of a
  driven-dissipative bose-hubbard system},
\newblock Science Advances \textbf{3}(12), e1701513 (2017),
\newblock \doi{10.1126/sciadv.1701513}.

\bibitem{Schnell2023}
A.~Schnell, L.-N. Wu, A.~Widera and A.~Eckardt,
\newblock \emph{Floquet-heating-induced bose condensation in a scarlike mode of
  an open driven optical-lattice system},
\newblock Phys. Rev. A \textbf{107}, L021301 (2023),
\newblock \doi{10.1103/PhysRevA.107.L021301}.

\bibitem{Wu2022}
L.-N. Wu and A.~Eckardt,
\newblock \emph{{Quantum engineering of a synthetic thermal bath for bosonic
  atoms in a one-dimensional optical lattice via Markovian feedback control}},
\newblock SciPost Phys. \textbf{13}, 059 (2022),
\newblock \doi{10.21468/SciPostPhys.13.3.059}.

\bibitem{Cheng2022}
Z.~Cheng and A.~C. Potter,
\newblock \emph{Matrix product operator approach to nonequilibrium floquet
  steady states},
\newblock Phys. Rev. B \textbf{106}, L220307 (2022),
\newblock \doi{10.1103/PhysRevB.106.L220307}.

\bibitem{Kirk1998}
D.~E. Kirk,
\newblock \emph{Optimal Control Theory. An Introduction},
\newblock Dover Publications, Inc., New York (1998).

\bibitem{Shapiro2003}
P.~Brumer and M.~Shapiro,
\newblock \emph{Principles of the Quantum Control of Molecular Processes},
\newblock John Wiley, New York (2003).

\bibitem{Brif2010}
C.~Brif, R.~Chakrabarti and H.~Rabitz,
\newblock \emph{{Control of quantum phenomena: past present and future}},
\newblock New Journal of Physics \textbf{12}(7), 075008 (2010).

\bibitem{Glaser2015}
S.~J. Glaser, U.~Boscain, T.~Calarco, C.~P. Koch, W.~K{\"o}ckenberger,
  R.~Kosloff, I.~Kuprov, B.~Luy, S.~Schirmer, T.~Schulte-Herbr{\"u}ggen,
  D.~Sugny and F.~K. Wilhelm,
\newblock \emph{Training schr{\"o}dinger's cat: quantum optimal control},
\newblock The European Physical Journal D \textbf{69}(12), 279 (2015),
\newblock \doi{10.1140/epjd/e2015-60464-1}.

\bibitem{Castro2018}
A.~Castro,
\newblock \emph{Optimal Control Theory for Electronic Structure Methods}, pp.
  1--21,
\newblock Springer International Publishing, Cham,
\newblock ISBN 978-3-319-42913-7,
\newblock \doi{10.1007/978-3-319-42913-7_4-1} (2018).

\bibitem{VargasHernandez2021}
R.~A. Vargas-Hernández, R.~T.~Q. Chen, K.~A. Jung and P.~Brumer,
\newblock \emph{Fully differentiable optimization protocols for non-equilibrium
  steady states},
\newblock New Journal of Physics \textbf{23}(12), 123006 (2021),
\newblock \doi{10.1088/1367-2630/ac395e}.

\bibitem{Krantz2003}
S.~Krantz and H.~Parks,
\newblock \emph{The implicit function theorem : History, theory, and
  applications / s.g. krantz, h.r. parks.}  (2003),
\newblock \doi{10.1007/978-1-4612-0059-8}.

\bibitem{Lindblad1976}
G.~Lindblad,
\newblock \emph{On the generators of quantum dynamical semigroups},
\newblock Communications in Mathematical Physics \textbf{48}(2), 119 (1976),
\newblock \doi{10.1007/BF01608499}.

\bibitem{Gorini1976}
V.~Gorini, A.~Kossakowski and E.~C.~G. Sudarshan,
\newblock \emph{Completely positive dynamical semigroups of n‐level systems},
\newblock Journal of Mathematical Physics \textbf{17}(5), 821 (1976),
\newblock \doi{10.1063/1.522979}.

\bibitem{Albash2012}
T.~Albash, S.~Boixo, D.~A. Lidar and P.~Zanardi,
\newblock \emph{Quantum adiabatic markovian master equations},
\newblock New Journal of Physics \textbf{14}(12), 123016 (2012).

\bibitem{Davies1978}
E.~B. Davies and H.~Spohn,
\newblock \emph{Open quantum systems with time-dependent hamiltonians and their
  linear response},
\newblock Journal of Statistical Physics \textbf{19}(5), 511 (1978),
\newblock \doi{10.1007/BF01011696}.

\bibitem{Lendi1986}
K.~Lendi,
\newblock \emph{Extension of quantum dynamical semigroup generators for open
  systems to time-dependent hamiltonians},
\newblock Phys. Rev. A \textbf{33}, 3358 (1986),
\newblock \doi{10.1103/PhysRevA.33.3358}.

\bibitem{Dann2018}
R.~Dann, A.~Levy and R.~Kosloff,
\newblock \emph{Time-dependent markovian quantum master equation},
\newblock Phys. Rev. A \textbf{98}, 052129 (2018),
\newblock \doi{10.1103/PhysRevA.98.052129}.

\bibitem{Alicki2007}
R.~Alicki and K.~Lendi,
\newblock \emph{Quantum dynamical semigroups and applications}, vol. 717,
\newblock Springer (2007).

\bibitem{Ikeda2020}
T.~N. Ikeda and M.~Sato,
\newblock \emph{General description for nonequilibrium steady states in
  periodically driven dissipative quantum systems},
\newblock Science Advances \textbf{6}(27), eabb4019 (2020),
\newblock \doi{10.1126/sciadv.abb4019}.

\bibitem{Kienzler2015}
D.~Kienzler, H.-Y. Lo, B.~Keitch, L.~de~Clercq, F.~Leupold, F.~Lindenfelser,
  M.~Marinelli, V.~Negnevitsky and J.~P. Home,
\newblock \emph{Quantum harmonic oscillator state synthesis by reservoir
  engineering},
\newblock Science \textbf{347}(6217), 53 (2015),
\newblock \doi{10.1126/science.1261033}.

\bibitem{Hartmann2017}
M.~Hartmann, D.~Poletti, M.~Ivanchenko, S.~Denisov and P.~Hänggi,
\newblock \emph{Asymptotic floquet states of open quantum systems: the role of
  interaction},
\newblock New Journal of Physics \textbf{19}(8), 083011 (2017),
\newblock \doi{10.1088/1367-2630/aa7ceb}.

\bibitem{Ikeda2021}
T.~N. Ikeda, K.~Chinzei and M.~Sato,
\newblock \emph{{Nonequilibrium steady states in the Floquet-Lindblad systems:
  van Vleck's high-frequency expansion approach}},
\newblock SciPost Phys. Core \textbf{4}, 033 (2021),
\newblock \doi{10.21468/SciPostPhysCore.4.4.033}.

\bibitem{Havel2003}
T.~F. Havel,
\newblock \emph{Robust procedures for converting among lindblad, kraus and
  matrix representations of quantum dynamical semigroups},
\newblock Journal of Mathematical Physics \textbf{44}(2), 534 (2003),
\newblock \doi{10.1063/1.1518555}.

\bibitem{qocttools}
A.~Castro,
\newblock \emph{qocttools},
\newblock Https://gitlab.com/acbarrigon/qocttools/.

\bibitem{VargasHernandez2020}
R.~A. Vargas-Hernández, R.~T.~Q. Chen, K.~A. Jung and P.~Brumer,
\newblock \emph{Inverse design of dissipative quantum steady-states with
  implicit differentiation} (2020), \eprint{2011.12808}.

\bibitem{Boltyanskii1956}
V.~G. Boltyanskiĭ, R.~V. Gamkrelidze and L.~S. Pontryagin,
\newblock \emph{On the theory of optimal processes. (russian)},
\newblock Dokl. Akad. Nauk SSSR (N.S.) \textbf{110}, 7 (1956).

\bibitem{Pontryagin1962}
L.~S. Pontryagin, V.~G. Boltyanskii, R.~V. Gamkrelidze and E.~F. Mishchenko,
\newblock \emph{The Mathematical Theory of Optimal Processes},
\newblock John Wiley \& Sons (1962).

\bibitem{PRXQuantum.2.030203}
U.~Boscain, M.~Sigalotti and D.~Sugny,
\newblock \emph{Introduction to the pontryagin maximum principle for quantum
  optimal control},
\newblock PRX Quantum \textbf{2}, 030203 (2021),
\newblock \doi{10.1103/PRXQuantum.2.030203}.

\bibitem{Cao2003}
Y.~Cao, S.~Li, L.~Petzold and R.~Serban,
\newblock \emph{Adjoint sensitivity analysis for differential-algebraic
  equations: The adjoint dae system and its numerical solution},
\newblock SIAM Journal on Scientific Computing \textbf{24}(3), 1076 (2003),
\newblock \doi{10.1137/S1064827501380630}.

\bibitem{Rondin2014}
L.~Rondin, J.-P. Tetienne, T.~Hingant, J.-F. Roch, P.~Maletinsky and
  V.~Jacques,
\newblock \emph{Magnetometry with nitrogen-vacancy defects in diamond},
\newblock Reports on Progress in Physics \textbf{77}(5), 056503 (2014),
\newblock \doi{10.1088/0034-4885/77/5/056503}.

\bibitem{Kraft1994}
D.~Kraft,
\newblock \emph{Algorithm 733: Tomp–fortran modules for optimal control
  calculations},
\newblock ACM Trans. Math. Softw. \textbf{20}(3), 262–281 (1994),
\newblock \doi{10.1145/192115.192124}.

\bibitem{nlopt}
S.~G. Johnson,
\newblock \emph{The nlopt nonlinear-optimization package},
\newblock Http://github.com/stevengj/nlopt.

\end{thebibliography}

\nolinenumbers

\end{document}